\documentclass[%
 reprint,
 amsmath,amssymb,
 aps,
pra,
]{revtex4-1}

\usepackage{amsthm}
\usepackage{graphicx}
\usepackage{dcolumn}
\usepackage{bm}
\usepackage{multirow} 
\usepackage{epstopdf}

\theoremstyle{remark}

\begin{document}

\makeatletter
\newcommand{\rmnum}[1]{\romannumeral #1}
\newcommand{\Rmnum}[1]{\expandafter\@slowromancap\romannumeral #1@}
\makeatother

\preprint{APS/123-QED}

\title{Quantum Conditional Random Field
}

\author{Yusen Wu$^{1,2}$}
\author{Chao-Hua Yu$^{1}$}
\author{Binbin Cai$^{1}$}
\author{Sujuan Qin$^{\dagger1}$}
\email{qsujuan@bupt.edu.cn}
\author{Fei Gao$^{1}$}
 \email{gaofei\_bupt@hotmail.com}
\author{Qiaoyan Wen$^{1}$}
\email{wqy@bupt.edu.cn}

\affiliation{%
 $^{1}$State Key Laboratory of Networking and Switching Technology, Beijing University of Posts and Telecommunications, Beijing, 100876, China\\
 $^{2}$State Key Laboratory of Cryptology, P.O. Box 5159, Beijing, 100878, China
}%

\date{\today}

\begin{abstract}
Conditional random field (CRF) is an important probabilistic machine learning model for labeling sequential data, which is widely utilized in natural language processing, bioinformatics and computer vision. However, training the CRF model is computationally intractable when large scale training samples are processed. Since little work has been done for labeling sequential data in the quantum settings, we in this paper construct a quantum CRF (QCRF) model by introducing well-defined Hamiltonians and measurements, and present a quantum algorithm to train this model. It is shown that the algorithm achieves an exponential speed-up over its classical counterpart. Furthermore, we also demonstrate that the QCRF model possesses higher Vapnik Chervonenkis dimension than the classical CRF model, which means QCRF is equipped with a higher learning ability.

\begin{description}
\item[PACS numbers]
03.67.Dd, 03.67.Hk
\end{description}
\end{abstract}

\pacs{Valid PACS appear here}
\maketitle


\section{\label{sec:level1}Introduction}

Quantum computing makes use of quantum mechanical phenomena, such as quantum superposition and quantum entanglement, to perform computing tasks on quantum systems, providing a new computing model fundamentally different from the classical computing \cite{Nielsen2002Quantum}. The most exciting thing about quantum computing is its ability to achieve significant speed-up over classical computing for solving certain problems, such as simulating quantum systems \cite{Bennett2001Optimal,Childs2012Hamiltonian,Low2017Optimal}, factoring large integer numbers \cite{Shor1997Polynomial}, and unstructured database searching \cite{Grover1996A}. In the past decade, the excitement has been brought into a newly emerging branch of quantum computing, quantum machine learning (QML), which is a interdisciplinary research field combining both quantum computing and machine learning \cite{Wittek2014Quantum}. Machine learning studies algorithms that assign a label (output) to each one of observations (inputs) by learning the model describing the relationship between the observation and the label. It mainly falls into two categories: supervised learning and unsupervised learning, depending on whether example observations and corresponding labels are provided or not. Since the pioneering quantum algorithm for linear systems of equations was proposed by Harrow et al. (HHL) \cite{Harrow2009Quantum}, a variety of quantum algorithms have been put forward to tackle various well-known machine learning problems, such as linear regression \cite{Schuld2016Prediction}, data classification \cite{Rebentrost2014Quantum}, clustering analysis \cite{Lloyd2013QuantumCluster}, principle component analysis \cite{Lloyd2013Quantum}, ridge regression \cite{Yu2017Quantum}, Toeplitz system \cite{lin2018Toeplitz}, polynomial approximation by gradient descent and Newton's method \cite{Rebentrost2016Quantum}, and so on. Fortunately, these quantum algorithms exhibit substantially significant speed-up over their counterparts. Recently, the study of QML has been extended to construct quantum neutral network models, such as quantum deep learning \cite{Wiebe2014Quantum}, quantum Boltzmann machines \cite{Kulchytskyy2016Quantum}, and quantum Hopfield neural network \cite{Rebentrost2018A}. Unfortunately, all of these algorithms and models only work for nonsequential data where observations (and labels) have no relationship with each other. However, sequential data arise in various fields including bioinformatics, speech recognition, and machine translation etc. \cite{Ghosh2009Markov}.

Conditional random field (CRF) is a probabilistic framework for labeling and segmenting sequential data, such as text sequences and gene sequences, where labels depend on other labels and observations \cite{Lafferty2001Conditional}. It plays an important role in machine learning, and has wide applications \cite{Fei2003Shallow,He2004Multiscale,Lafferty2001Conditional,Settles2004Biomedical} in the fields of natural language processing (NLP), bioinformatics, and computer vision. Given $n$ sequential observations, $\mathbf{x}=(\mathbf{x}_1,\cdots,\mathbf{x}_n)$, and their corresponding labels, $\mathbf{y}=(y_1,\cdots,y_n)$, CRF aims to model the conditional probability distribution $P(\mathbf{y}|\mathbf{x})$. In the model, some eigen-functions $\{f_k\}_{k=1}^{K}$ are introduced to describe the inner relationships within the observation sequence $\mathbf{x}$ and the label sequence $\mathbf{y}$. Meanwhile the eigen-functions also describe the relationship between them. The CRF model is parameterized by the coefficients $w_k$ of the eigen-functions $f_k$. A simple but motivating example from NLP is part-of-speech tagging, in which $\mathbf{x}_s$ denotes the word of the sentence $\mathbf{x}$ at position $s$, and $y_s$ is its corresponding part-of-speech tag. Training the CRF model generally uses the gradient decent method to obtain the parameters $w_k$ in an iterative way, and the time complexity grows exponentially with $n$ \cite{Lafferty2001Conditional}. It means that training a CRF model will be computationally intractable when $n$ is large. After the parameters $w_k$ are obtained, the model can be used to predict the label for any new observation via the efficient Viterbi algorithm \cite{Viterbi2003Viterbi}.

In this paper, we explore how to model and train CRF on quantum settings. Specifically, by introducing two well-defined Hamiltonians both encoding the parameters $w_k$, as well as two well-defined measurement operators both encoding the eigen-functions $f_k$, we construct a quantum CRF (QCRF) model where the conditional probability distribution $P(\mathbf{y}|\mathbf{x})$ is derived by simple linear algebra operations on the exponents of the Hamiltonians and the measurement operators. We also present a quantum algorithm for training the model, which uses state-of-the-art Hamiltonian simulation \cite{Low2017Optimal} to obtain the classical information of the parameters $w_k$. It is shown that the time complexity of our algorithm grows polynomially with $n$ (i.e., the number of example observation-label couples for training), exponentially improving the dependence on $n$ compared with the classical training algorithm mentioned above.
Furthermore, we also compare the QCRF and the classical CRF from the perspective of computational learning theory. We show that, our QCRF model has a much higher Vpanik-Chervonenkis (VC) dimension \cite{Vapnik1971On} than the classical CRF, demonstrating that QCRF significantly improves the data learning ability over the classical CRF.

\section{Review of classical CRF}
\label{sec:2}
In this section, we first review the definition of CRF model and the methodology to train it.
\subsection{Definition of CRF model}

Suppose $X$ and $Y$ are random variables, and $G(V,E)$ is an undirected graph such that $Y=(Y_{v}),v\in V$. The vertex set $V$ represents random variables and edge set $E$ stands for the dependency relationships between random variables. Then $(X,Y)$ is a CRF when the random variables $Y_v$, conditioned on $X$, obey the Markov property with respect to the graph $G$, i.e. $P(Y_{v_1}|X,Y_{v_2},v_1\neq v_2)=P(Y_{v_1}|X,Y_{v_2},v_1\sim v_2)$ is satisfied for every node $v_1$, where $v_1\sim v_2$ means that $v_1$ and $v_2$ are neighbors in $G$. $Y_{v_1}$ and $Y_{v_2}$ are  corresponding random variables of vertices $v_1, v_2$. Theoretically, we can construct a structure that fully models the graph $G$ in any arbitrary complexity, and the most commonly used Chain-structured CRF is defined as
\begin{eqnarray}
\label{Eq:1}
P(\mathbf{y|x})=\frac{1}{Z}\exp\{\sum_{i=1}^{n}\sum_{k=1}^{K}w_{k}f_{k}(x_i,y_i)\}.
\end{eqnarray}
Here $x_i$ and $y_i$ respectively denote the $i$-th observation and its corresponding label, $f_{k}(x_i,y_i)$ takes value in $\{-1,1\}$ with Boltzmann weight $w_k$, and $Z$ is the normalization factor.  If $y_i$ is assigned to $x_i$ with a high probability, $f_{k}(x_i,y_i)$ returns value $1$, otherwise returns $-1$.

Similar to most supervised learning models, CRF also involves two phases, namely training phase and predicting phase. Specifically, the training phase analyzes the training data to construct the most appropriate mapping $P(\mathbf{y}|\mathbf{x})$, in which $\mathbf{x}=(x_i)_{i=1}^{n}$ and $\mathbf{y}=(y_i)_{i=1}^{n}$ denote the observation sequence and label sequence, respectively. The predicting phase aims at computing the most probable label sequence $\mathbf{y_0}$ for new observation sequence $\mathbf{x_0}$ with the help of $P(\mathbf{y}|\mathbf{x})$ obtained in the training phase.
\subsection{Training CRF}

 Given large scale of observation sequence $\mathbf{x}$ and corresponding label sequence $\mathbf{y}$, CRF training phase suffices to construct the inference model $P(\mathbf{y|x})$ on the graph, i.e. finding the Boltzmann weights $w=(w_1,..,w_K)$. One approach to find the appropriate $w$ is the minimum-likelihood method based upon the observable $\mathbf{x}$ and label $\mathbf{y}$. If the training process is successful, the model joint distribution $P_{x,y}$ has enough resemblance to the priori data joint distribution $P_{x,y}^{data}$.
To describe this approaching, we introduce the log-likely hood function $L$, whose minimum point corresponds to the appropriate $w$, or the appropriate relationship. Suppose the average negative log-likely hood function $L$ is defined as \cite{Lafferty2001Conditional}
\begin{eqnarray}
\label{Eq:2}
L=-\sum\limits_{x}P_{x}^{data}\sum\limits_{y}P^{data}(y|x)\log P(y|x)
\end{eqnarray}
\begin{eqnarray}
\label{Eq:4}
 =-\sum\limits_{x,y}P_{x,y}^{data}\log\frac{e^{E(x,y)}}{\sum\limits_{y^{*}}e^{E(x,y^{*})}}.
\end{eqnarray}
The summation on $y^{*}$ indicates traversing all the possible label sequences with the fixed observation data $\mathbf{x}$, and the potential function $E(x,y)$ is in the form of
\begin{eqnarray}
\label{Eq:5}
E(\mathbf{x,y})=\sum\limits_{i=1}^{n}\sum\limits_{k=1}^{K}w_{k}f_{k}(x_i,y_i). \end{eqnarray}
To determine the Boltzmann weights $w$, we need to minimize the function $L$ with the help of optimization formulas, e.g. Newton method, BFGS method and Gradient descent method. It is interesting to note that these methods all depend on calculating the gradient of $L$. In each iteration, the parameter $w$ is updated by a selected step in the direction opposite to the gradient: $\Delta w=-\eta \frac{\partial L}{\partial w}$, where $\eta$ is the step length and the gradient $\frac{\partial L}{\partial w}$ is expressed as
\begin{eqnarray}
\label{Eq:5}
 \frac{\partial L}{\partial w_k}= -\sum\limits_{x,y}P_{x,y}^{data}(\frac{\frac{\partial}{\partial w_k}e^{E(x,y)}}{e^{E(x,y)}}-\frac{\sum\limits_{y^{*}}\frac{\partial}{\partial w_k}e^{E(x,y^{*})}}{\sum\limits_{y^{*}}e^{E(x,y^{*})}})
\end{eqnarray}
\begin{eqnarray}
\label{Eq:5}
=-\sum\limits_{x,y}P_{x,y}^{data}(\langle e^{E}\rangle_{X,Y}^{C,k}-\langle e^{E}\rangle_{X}^{C,k}).
\end{eqnarray}
Noting that the gradient function $\frac{\partial L}{\partial w_k}$ has two terms, in which the first term $\langle e^{E}\rangle_{X,Y}^{C,k}$ is clamped by the training data $\{\mathbf{x,y}\}$, and the second term $\langle e^{E}\rangle_{X}^{C,k}$ is only clamped by the observable set $\mathbf{x}$. Updating gradient achieves exponential complexity in classical computing, since the second term traverses all the possible $y^{*}$ leading to $\mathcal{O}(\|Q\|^{n})$ possibilities, and each possibility $\frac{\partial}{\partial w_k}e^{E(x,y^{*})}$ has $\mathcal{O}(n)$ terms, which is intractable facing large scale $n$.

In the following, we will first give the theory of QCRF model and then present a quantum algorithm for training the model that is prominently exponentially faster than the classical CRF training algorithm.

\section{QCRF model}
\subsection{Fundamental theory of QCRF}
 Since classical CRF model is proposed according to the classical conditional entropy, we propose the fundamental theory of QCRF model based on the principle of Quantum Conditional Entropy \cite{Nielsen2002Quantum}. For the QCRF model, the quantum conditional entropy $S(\mathbf{y|x})$ is applied, which is defined as
\begin{eqnarray}
\label{Eq:6}
S(\mathbf{y|x})=S(\rho _{X,Y})-S(\rho _{X}).
\end{eqnarray}
The density operator $\rho _{X,Y}$ encodes the joint distribution of $P(x,y)$ in its amplitude, which can be decomposed on its spectrum with the probability $P(x,y)$, i.e., $\rho _{X,Y}=\sum\limits_{x,y}P(x,y)|x,y\rangle\langle x,y|$. And the density operator $\rho_ {X}$ indicates the marginal distribution $P(x)$ of random variable $X$, similarly, we can also decompose $\rho _{X}$ into the form of $\rho _{X}=\sum\limits_{x}P(x)|x\rangle\langle x|$. Thus the quantum conditional entropy $S(\mathbf{y|x})$ can be expressed as
\begin{eqnarray}
\label{Eq:7}
S(\mathbf{y|x})=-\sum\limits_{x,y}P(x,y)\log(P(y|x)).
\end{eqnarray}
The kernel idea behind the QCRF model is to find the model $P^{*}(\mathbf{y|x})$. Applying quantum conditional entropy, we obtain the form of objective function:
\begin{eqnarray}
\label{Eq:8}
P^{*}(\mathbf{y|x})=\arg \, \max_{P(\mathbf{y|x})}S(\mathbf{y|x}).
\end{eqnarray}
The model $P^{*}(\mathbf{y|x})$ has the largest possible quantum conditional entropy, which is still consistent with the information from the training material. Finding the quantum conditional probabilistic model under some constraints can be formulated as an optimization problem. We mainly take two constraints into consideration. The first constraint depends on the training material, which requires the model distribution should be close to the priori empirical distribution. That means, for each eigen-function $f_k$, its expected value on the empirical distribution must be equal to its expected value on the model distribution. The empirical distribution of $f_k$ is obtained by simply counting how often the different values of the variable occur in the training data. Introducing the Lagrange multiplier $w_k$, we obtain the priori distribution:
\begin{eqnarray}
\label{Eq:10}
E_{C}(f_1,...,f_K)=\frac{1}{N}\sum\limits_{k=1}^{K}\sum\limits_{(x,y)}w_k f_k(x,y).
\end{eqnarray}
In contrast to the priori empirical distribution, the quantum model distribution of $f_k$ is formulated as
\begin{eqnarray}
\label{Eq:9}
E_{Q}(f_1,...,f_K)=Tr(\Lambda _{X,Y}(\rho _{X,Y}H^{(0)})).
\end{eqnarray}
The notation $Tr(\cdot)$ means the trace of a matrix. Hamiltonian $H^{(0)}$ is composed by all the base of Hilbert space $\mathcal{H}$, where $\mathcal{H}=\{(\sum\limits_i\sum\limits_k w_{k}f_{k,i})|u_j\rangle |f_{k,i}\in\{-1,1\}\}$ and $|u_j\rangle$ denotes a set of base spanning the Hilbert space $\mathcal{H}$. Density matrix $\rho_{X,Y}$ encodes the joint distribution $P(x,y)$ corresponding to every combination of features $f_{k,i}$. And $\Lambda _{X,Y}=\sum\limits_{(x,y)}\Lambda(x,y)|x,y\rangle\langle x,y|$ is the measurement operator limiting the Hamiltonian only to the clamped $X,Y$ provided by the training data set. The parameter $\Lambda(x,y)=1$ if and only if $x=X, y=Y$, otherwise $\Lambda(x,y)=0$.\\
The second constraint claims the normalization condition:
$\sum\limits_{y^{*}}P(y^{*}|x)=1$. After introducing another Lagrange parameter $\lambda$ on the normalization condition, the Lagrange formula function can be interpreted as:
\begin{multline}
\label{Eq:11}
G(P(\mathbf{y|x}))=S(\mathbf{y|x})+(E_{Q}(f_1,...,f_K)\\
-E_{C}(f_1,...,f_K))+\lambda (\sum\limits_{y^{*}}P(y^{*}|x)-1).
\end{multline}
Equating the partial derivative $\frac{\partial G(P(y|x))}{\partial P(y|x)}$ to $0$ and solving by $P(y|x)$, we obtain
\begin{eqnarray}
\label{Eq:13}
P(\mathbf{y|x})=Tr(\Lambda_{X,Y}e^{H^{(0)}})\exp(\lambda/P(x)-1).
\end{eqnarray}
It is worthy to note that $\sum\limits_{y^{*}}P(y^{*}|x)=1$, then we have
\begin{eqnarray}
\label{Eq:14}
\sum\limits_{y^{*}}Tr(\Lambda_{X,Y}e^{H^{(0)}})\exp(\lambda/P(x)-1)=1.
\end{eqnarray}
To simplify Eq. (14), we introduce another Hamiltonian $H^{(n)}=I_Q^{\otimes n}\otimes H^{(0)}$ and corresponding measurement $\Lambda _X=\sum\limits_x\Lambda(x)|x,y\rangle\langle x,y|$, which traverse all the possible $y^{*}$, then we have
\begin{eqnarray}
\label{Eq:15}
\sum\limits_{y^{*}}Tr(\Lambda_{X,Y}e^{H^{(0)}})=Tr(\Lambda_{X}e^{H^{(n)}}). \end{eqnarray}
The parameter $\Lambda(x)=1$ if and only if $x=X$, otherwise $\Lambda(x)$ equals to $0$. Combining Eqs. (13), (14) and (15),
the QCRF model can be formulated as
\begin{eqnarray}
\label{Eq:15}
P(\mathbf{y|x})=\frac{Tr(\Lambda_{X,Y}e^{H^{(0)}})}{Tr(\Lambda_{X}e^{H^{(n)}})}.
\end{eqnarray}
This is the kernel expression of QCRF model. In the following,  we illustrate a simple method to construct the concrete Hamiltonians $H^{(0)}$, $H^{(n)}$ and corresponding measurement $\Lambda_{X,Y}$, $\Lambda_{X}$.
\subsection{The construction of Hamiltonian and measurement}
 One of the kernel target of QCRF model aims at constructing the special designed Hamiltonian to represent the potential function $E(x,y)$ via replacing the classical bits with quantum bits. The potential function $E(x,y)$ not only reflects the statistical property, but also describes the matching degree between the observation variable $\mathbf{x}$ and label sequence $\mathbf{y}$ by eigen-function $f_k(x_i,y_i)$. To simulate the statistical property and the matching degree of the variables, we utilize the Pauli Z analogue density operator $\sigma _{k,i}^{z,l}$ to reflect the empirical characteristics of data.
Define the Pauli Z analogue operator $\sigma _{k,i}^{z,l}$ as follows:
\begin{eqnarray}
\label{Eq:16}
\sigma_{k,i}^{z,l}=I_Q^{\otimes l}\otimes I_{2}^{\otimes(ki-1)}\otimes \sigma^z\otimes I_2^{\otimes(Kn-ki)}
\end{eqnarray}
The tag parameter $l$ on the $\sigma_{k,i}^{z,l}$'s shoulder controls the number of identity gate $I_Q$. The notation $\otimes$ means Kronecker product. And the operator $\sigma^{z}$ indicates the Pauli Z operator \cite{Nielsen2002Quantum}.
Every element in the expression $\sigma_{k,i}^{z,l}$ is an identity operator except the $(ki+l)$ -th $\sigma^{z}$ operator. This construction promises $\sigma_{k,i}^{z,l}$ emerging in form of diagonal operator, which encodes equivalent statistical distribution on its diagonal. Specifically, Hamiltonian $H^{(0)}$ and $H^{(n)}$ can be expressed as
\begin{eqnarray}
\label{Eq:16}
H^{(0)}=\sum\limits_{i=1}^{n}\sum\limits_{k=1}^{K}w_{k}\sigma _{k,i}^{z,0}
\end{eqnarray}and
\begin{eqnarray}
\label{Eq:16}
H^{(n)}=\sum\limits_{i=1}^{n}\sum\limits_{k=1}^{K}w_{k}\sigma _{k,i}^{z,n}
\end{eqnarray}
respectively.
The parameter $w_k$ also indicates the Boltzmann weight expressed in the potential function $E(x,y)$.
Different with the classical CRF model describes the data patterns via the binary eigen-function $f_k(x,y)$, QCRF model depends on the natural physical mechanism of particle topspin and downspin.

To construct the measurement operator, we first define the subspace $S_0$ and $S_n$ as follows:
$S_{0}=e^{E(x,y)}|\varphi_i\rangle$, where $ |\varphi_i\rangle$ is one of the base according to the index $i\in \{1,2,...,2^{nK}\}$, and $S_{n}=span\{e^{E(x,y^{(i)})}|\psi_i\rangle \\|i=1,2,\cdots,\|Q\|^{n}\}$.
Thus $S_l\subset e^{H^{(l)}}$, $l=0$ or $n$, whose trace norm satisfies $\|S_0\|_{tr}=e^{E(x,y)}$ and $\|S_n\|_{tr}=\sum\limits_{y^{*}}e^{E(x,y^{*})}$.
Since the Hilbert space $e^{H^{(l)}}$ is separable, we can decompose the space $e^{H^{(l)}}$ into the form of $e^{H^{(0)}}=S_0\oplus S_0^{\perp}$ and $e^{H^{(n)}}=S_n\oplus S_n^{\perp}$, in which $S_{l}^{\perp}$ is the orthogonal complementary space of $S_l$ defined on $e^{H^{(l)}}$, $l=0$ or $n$. For a subspace, $S_{l}$ is a diagonal matrix with its elements clamped by the variables $x,y$. To extract this subspace, we should design specific projection operators $\Lambda$, which projects the whole space $e^{H^{(l)}}$  only onto the subspace $S_l$.

Extracting subspace $S_0$ from whole space $e^{H^{(0)}}$ depends on the information of training data $x,y$. The properties of training data are reflected by the eigen-function series $f_k(x_i,y_i)$, therefore, eigen-function can provide all the information of $S_0$. Specifically, the corresponding projection operator $\Lambda_{X,Y}$ is expressed as:
\begin{eqnarray}
\label{Eq:20}
\Lambda_{X,Y}=\prod\limits_{i=1}^{n}\prod\limits_{k=1}^{K}\frac{1}{2}(I+f_{k}(x_i,y_i)\sigma_{k,i}^{z,0}).
\end{eqnarray}
The construction of $\frac{1}{2}(I+f_{ki}\sigma_{k,i}^{z,0})$ promises its entries are only onto the diagonal with the eigen-value of $1$ or $0$. The operator $\Lambda_{X,Y}$ is designed to perform on $Kn$ qubits. Given a quantum system consisting of $Kn$ qubits, $\Lambda_{X,Y}$ can be utilized to test whether the quantum system collapses on the state $\otimes|\frac{1-f_k(x_i,y_i)}{2}\rangle$ or not.
With the help of projection operator $\Lambda_{X,Y}$, we obtain the subspace $S_0$:
\begin{eqnarray}
\label{Eq:21}
S_0=\Lambda_{X,Y}(e^{H^{(0)}}).
\end{eqnarray}
The subspace $S_n$ can also be extracted in a similar method. The fundamental difference just lies in the fact that $S_n$ traverses all the possible $y^{*}$. We need to find out every $y^{*}$ with the help of eigen-function $f_k(x_i,y_i^{*})$. As a result, the project operator $\Lambda _X$ can be expressed into the form of:
\begin{eqnarray}
\label{Eq:22}
\Lambda_{X}=\prod\limits_{i=1}^{n}\prod\limits_{k=1}^{K}\frac{1}{2}\left(I+\sum\limits_{j=1}^{\|Q\|}f_k(x_i,y_i^{(j)})|j\rangle\langle j|\otimes \sigma_{k,i}^{z,n-1}\right).\nonumber\\
\end{eqnarray}
The notation $y_i^{(j)}$ indicates all the circumstances for $y_i$. The measurement operator $\Lambda_{X}$ is established on a $(\log\|Q\|+K)n$ qubits quantum system, which can be utilized testing whether the quantum system collapses on the state $|j\rangle\otimes|\frac{1-f_1(x_i,y_i^{(j)})}{2}\rangle\otimes...\otimes|\frac{1-f_K(x_i,y_i^{(j)})}{2}\rangle$ or not. We can extract the subspace $S_n$  as follows:
\begin{eqnarray}
\label{Eq:23}
S_n=\Lambda_{X}(e^{H^{(n)}}).
\end{eqnarray}
Noting that the trace norm of subspace $S_0$ and $S_n$  respectively equal to $e^{E(x,y)}$ and marginal distribution $\sum\limits_{y^*}e^{E(x,y^*)}$, then we can obtain the relationships
\begin{eqnarray}
\label{Eq:24}
Tr(\Lambda_{X,Y}e^{H^{(0)}})=e^{E(x,y)}
\end{eqnarray}
and
\begin{eqnarray}
\label{Eq:24}
Tr(\Lambda_{X}e^{H^{(n)}})=\sum\limits_{y^{*}}e^{E(x,y^{*})},
\end{eqnarray}
which build up a bridge between the quantum model and classical information. Up to now, we utilize the Hamiltonian $H^{(0)}$ and $H^{(n)}$ representing the gradient function $\frac{\partial L}{\partial w_k}$ under the quantum model:
\begin{eqnarray}
\label{Eq:25}
\frac{\partial L}{\partial w_k}=-\sum\limits_{x,y}P_{x,y}^{data}(\langle e^{H^{(0)}}\rangle _{X,Y}^{Q,k}-\langle e^{H^{(n)}}\rangle _{X}^{Q,k}),
\end{eqnarray}
where
\begin{eqnarray}
\label{Eq:25}
\langle e^{H^{(0)}}\rangle _{X,Y}^{Q,k}=\frac{Tr(\Lambda_{X,Y}\frac{\partial}{\partial w_k}e^{H^{(0)}})}{Tr(\Lambda_{X,Y}e^{H^{(0)}})}
\end{eqnarray} and
\begin{eqnarray}
\label{Eq:25}
\langle e^{H^{(n)}}\rangle _{X}^{Q,k}=\frac{Tr(\Lambda_{X}\frac{\partial}{\partial w_k}e^{H^{(n)}})}{Tr(\Lambda_{X}e^{H^{(n)}})}.
\end{eqnarray}
The Eq. (25) is a pivotal component utilized in QCRF training process, which encapsulates static property of Hamiltonian $H^{(l)}$ and the information provided by training data. Besides, the joint probabilities $P_{x,y}^{data}$ are given priori. Thus we can only concentrate on computing the terms $\langle e^{H^{(0)}}\rangle _{X,Y}^{Q,k}$ and $\langle e^{H^{(n)}}\rangle _{X}^{Q,k}$. Further training the QCRF model, or finding the Boltzmann weights $w$, also relies on this expression.
\section{Quantum algorithm for training QCRF}
 The above section proposes the fundamental theory of QCRF model. In this section, we present a quantum algorithm for training the QCRF model, then analyze its time complexity. Finally, a numerical simulations are performed on both CRF and QCRF models. Surprisingly, the results show that, in contrast to CRF model, our QCRF model requires significantly fewer iterations to achieve the same error rate.
\subsection{Algorithm}
The kernel target of our quantum algorithm aims at estimating the average terms $\langle e^{H^{(0)}}\rangle _{X,Y}^{Q,k}$ and $\langle e^{H^{(n)}}\rangle _{X}^{Q,k}$, as shown in Eqs. (27) and (28), thereby the gradient function $\partial L/\partial w_k$ can be computed efficiently.

\begin{figure*}
  \begin{center}
  \includegraphics[width=0.9\textwidth]{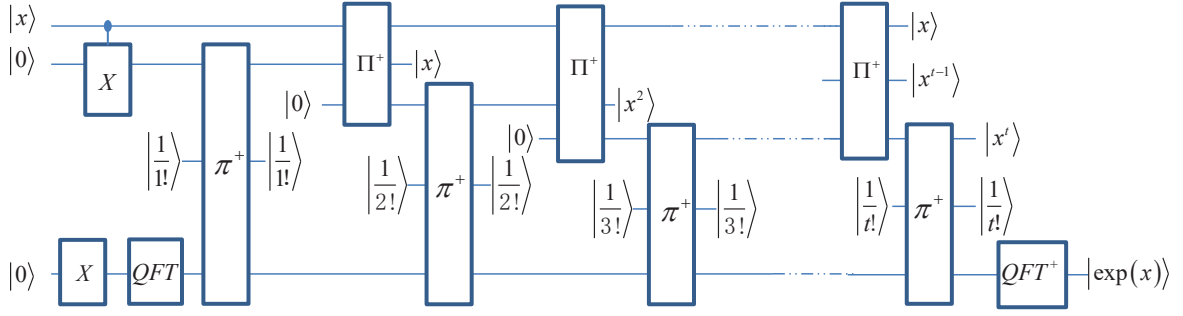}

  \caption{EXP gate. We introduce the quantum multiply gate for real inputs and outputs , which can realize the following transformation,
$\Pi_{m,n}^{+}|a\rangle|b\rangle|c\rangle=|a\rangle|b\rangle|c+ab\rangle$,
where $m,n$ denoted the  number of digits of $a$ and $b$ respectively \cite{Zhou2017Quantum}. This quantum multiply gate can be decomposed into the form:
$\Pi_{m,n}^{+}=(I\otimes I\otimes QFT^{\dagger})\pi_{m,n}^{+}(I\otimes I\otimes QFT)$.
$\pi_{m,n}^{+}$ is the intermediate multiply adder, which achieves the transformation
$\pi_{m,n}^{+}|a\rangle|b\rangle|\varphi(c)\rangle=|a\rangle|b\rangle|\varphi(c+ab)\rangle,$
with $|\varphi(c)\rangle:=QFT|c\rangle$. Utilizing the gate $\pi_{m,n}^{+}$ and $\Pi_{m,n}^{+}$, we achieve the EXP gate.
}\label{Figure1}
\end{center}
\end{figure*}

\begin{figure}[t]
\begin{center}
\includegraphics[width=0.45\textwidth]{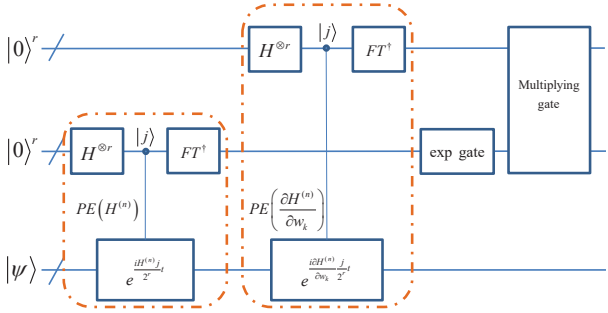}\\

\end{center}
 \caption{The quantum circuit of the 3-5 steps in our quantum algorithm. In detail, the $\exp$ gate implements $\exp|a\rangle=|e^{a}\rangle$, and the Multiplying gate performs $Muti|a\rangle|b\rangle|0\rangle=|a\rangle|b\rangle|ab\rangle$. }\label{Figure2}
\label{fig:long}
\label{fig:onecol}
\end{figure}

Evidently, estimating $\langle e^{H^{(0)}}\rangle _{X,Y}^{Q,k}$ and $\langle e^{H^{(n)}}\rangle _{X}^{Q,k}$ requires us to estimate the four terms $Tr(\Lambda_{X}\frac{\partial}{\partial w_k}e^{H^{(n)}})$, $Tr(\Lambda_{X}e^{H^{(n)}}), Tr(\Lambda_{X,Y}\frac{\partial}{\partial w_k}e^{H^{(0)}})$, and $Tr(\Lambda_{X,Y}e^{H^{(0)}})$. Since these four terms have almost the same form, we just concentrate on estimating the relatively complicated term $Tr(\Lambda_{X}\frac{\partial}{\partial w_k}e^{H^{(n)}})$, which is mathematically equal to $Tr(\Lambda_{X}e^{H^{(n)}}\frac{\partial}{\partial w_k}H^{(n)})$. From Eqs. (19) and (22), it is evident that $\Lambda_X$, $e^{H^{(n)}}$ and $\frac{\partial}{\partial w_k}H^{(n)}$ have the same eigenvectors we denote as $\{|\psi_k\rangle\}_{k=1}^{D}$, where $D=\|Q\|^n2^{nK}$ is the dimension of these three operators. To estimate $Tr(\Lambda_{X}\frac{\partial}{\partial w_k}e^{H^{(n)}})$, our algorithm will first generates $m$ copies of the state
\begin{eqnarray}
\label{Eq:31}
|\phi\rangle=\frac{1}{\sqrt{\sum\limits_k\|\lambda_k\|}}\sum\limits_{k=1}^{D}\sqrt{\lambda_k}|\psi_k\rangle,
\end{eqnarray}
which are denoted by $|\phi_j\rangle, j=1,2,...,m$.
Then we perform measurement $\Lambda_X$ on each one to estimate the desired average, i.e.,
\begin{eqnarray}
\label{Eq:28}
Tr(\Lambda_{X}\frac{\partial}{\partial w_k}e^{H^{(n)}})= \frac{1}{m}\sum\limits_{j=1}^{m}\langle \phi_j|\Lambda_X|\phi_j\rangle.
\end{eqnarray}
The whole quantum  algorithm can be summarized as follows.\\
\textbf{Algorithm:} Training QCRF model\\
\textbf{Input:}Boltzmann weight $w=(w_1,w_2,...,w_K)$, training data set $(\mathbf{x,y})$, Hamiltonian $H^{(0)}$, $H^{(n)}$ and measurement operators $\Lambda_{X,Y}$, $\Lambda_{X}$.\\
\textbf{Output:}The estimation of gradient function $\frac{\partial L}{\partial w_k}$
\begin{enumerate}
  \item Initial state: $|0\rangle_{(n(\log Q+K)}|0\rangle_{r}|0\rangle_{r}$. The first system consists of $n(\log Q+K)$ qubits to present the $\|Q\|^n2^{nK}$ dimension superposition. The second and third system are encoded with $r$ qubits of precision.
  \item Perform Hadamard operator $H$ on the first system implementing the superposition
\begin{eqnarray}
\label{Eq:30}
H|0\rangle=\frac{1}{\sqrt{D}}\sum\limits_{k=1}^{D}|\psi_k\rangle|0\rangle|0\rangle,
\end{eqnarray}
  where $D$ is the dimension of Hamiltonian $\frac{\partial}{\partial w_k}e^{H^{(n)}}$.
  \item Noting that the Hamiltonian $H^{(n)}$ can be decomposed on the computational basis $|\psi_i\rangle$. Then perform the phase estimation $PE(H^{(n)})$ on the first and second register. The system achieves to
\begin{eqnarray}
\label{Eq:30}
\frac{1}{\sqrt{D}}\sum\limits_{k=1}^{D}|\psi_k\rangle|E_k\rangle|0\rangle,
\end{eqnarray}
where $E_k$ indicates the eigen-value of Hamiltonian $H^{(n)}$ corresponding to $|\psi_k\rangle$. Implementing the phase estimation $PE(H^{(n)})$ depends on Hamiltonian simulation, which achieves the controlled operator $e^{\frac{-iH^{(n)}jt}{2^r}}$.

  \item Then perform the phase estimation $PE(\frac{\partial}{\partial w_k}H^{(n)})$ on the third register, and the system obtains the phase $\mu_i$ of the matrix $\frac{\partial}{\partial w_k}H^{(n)}$ in the third register
\begin{eqnarray}
\label{Eq:31}
\frac{1}{\sqrt{D}}\sum\limits_{k=1}^{D}|\psi_k\rangle|E_k\rangle|\mu_k\rangle.
\end{eqnarray}

  \item Invoking exponential gate on the second register, whose quantum circuit is illustrated in Fig. \ref{Figure1}. Then executing the multiplication operator onto the second and third registers. After that, undo the phase estimation $PE(\frac{\partial}{\partial w_k}H^{(n)})$. The system becomes to
\begin{eqnarray}
\label{Eq:31}
\frac{1}{\sqrt{D}}\sum\limits_{k=1}^{D}|\psi_k\rangle|\mu_k e^{E_k}\rangle|0\rangle.
\end{eqnarray}
Fig. \ref{Figure2} illustrates the quantum circuit of the steps 3-5.

  \item Denote $\mu_k e^{E_k}$ as $\lambda_k$, and apply the controlled rotation onto the third system, then the system becomes to
\begin{eqnarray}
\label{Eq:31}
\frac{1}{\sqrt{D}}\sum\limits_{k=1}^{D}|\psi_k\rangle|\lambda_k\rangle(q_k|0\rangle+\sqrt{1-q_k^2}|1\rangle),
\end{eqnarray}
where $q_k=C\sqrt{\lambda_k}$ is a normalisation factor.

\item Measuring the last register to see the outcome $|0\rangle$ and discarding it. Then undo the phase estimation of the second system, we have the state
\begin{eqnarray}
\label{Eq:31}
|\phi\rangle=\frac{1}{\sqrt{\sum\limits_k\|\lambda_k\|}}\sum\limits_{k=1}^{D}\sqrt{\lambda_k}|\psi_k\rangle,
\end{eqnarray}
with probability $P(0)=C^2\sum\limits_k\|\lambda_k\|/D$.
The lower bound of probability $P(0)$ can be estimated as $\Omega(C^2(1-nK\max\|w\|/D))$. Then choosing the parameter $C=\sqrt{D(1-\varepsilon)/(D-nK\max\|w\|)}$ promises the system can measure the $|0\rangle$ with a relative high probability $1-\varepsilon$. Furthermore, we can also utilize the Amplitude Amplification \cite{Brassard2012Quantum} method to enhance $P(0)$. This state $|\phi\rangle$ is then moved over and stored in a quantum memory. Then reinitializing  quantum computer and repeating steps 1-7 for $m$ times, and we obtain $m$ states $|\phi_1\rangle,...,|\phi_{m}\rangle$ storing in the quantum memory.
  \item
  Finally, we can estimate the term $Tr(\Lambda_{X}\frac{\partial}{\partial w_k}e^{H^{(n)}})$ by measuring the observable $\Lambda_{X}$ with the states $|\phi_1\rangle,...,|\phi_m\rangle$:
\begin{eqnarray}
\label{Eq:31}
Tr(\Lambda_{X}\frac{\partial e^{H^{(n)}}}{\partial w_k})=\frac{P(0)D}{mC^2}\sum\limits_{i=1}^{m}\langle \phi_i|\Lambda_X|\phi_i\rangle+\varepsilon_{m},
\end{eqnarray}
where $\varepsilon_{m}$ is the measurement error. Noting that measurement operator $\Lambda_X$ is constructed in the form of continuous product of simple operators, and this construction implies our measurement can be easily achieved. The probability $P_{\Lambda_X}$ reflects the measurement results on $|j\rangle\otimes|\frac{1-f_1(x_i,y_i^{(j)})}{2}\rangle\otimes...\otimes|\frac{1-f_K(x_i,y_i^{(j)})}{2}\rangle$ . Repeat the measurement $\mathcal{O}(m)$ times, we can acquire the statistical value of $\langle\phi|\Lambda_X|\phi\rangle$.
Furthermore we can also utilize the similar method to estimate the numerical value of $Tr(\Lambda_{X}e^{H^{(n)}}), Tr(\Lambda_{X,Y}\frac{\partial}{\partial w_k}e^{H^{(0)}})$, and $Tr(\Lambda_{X,Y}e^{H^{(0)}})$. Then the terms $\langle e^{H^{(0)}}\rangle _{X,Y}^{Q,k}$ and $\langle e^{H^{(n)}}\rangle _{X}^{Q,k}$ can be computed efficiently. Finally, the gradient function $\frac{\partial L}{\partial w_k}$ gains.
\end{enumerate}

\subsection{Complexity analysis}
 We continue with a discussion of the run time of our quantum algorithm. First, the Hadamard operator $H$ performs on $(\log Q+K)n$ qubits, thus Hadamard operator takes $\mathcal{O}(n)$ time to generate a superposition. Then, the Hamiltonian simulating is one of the preliminary theories of our quantum algorithm. The cost of simulating the time evolution operator $e^{-iHt}$ depends on several factors: the number of system gates, evolution time $t$, target error $\varepsilon_{si}$, and how information on the Hamiltonian $H$ is made available. In detail, Qubitization method \cite{Low2016Hamiltonian}, which achieves the simulation optimal bound in theory, can be made fully constructive with an approach for implementing some signal states $|G\rangle=\sum\limits_{k,i}\sqrt{\frac{w_k}{\alpha}}|k\rangle|i\rangle$ and signal operators $U=\sum\limits_{k,i}|i,k\rangle\langle i,k|\otimes\sigma _{k,i}^{z,l}$ that encode Hamiltonian $H^{(l)}$. It is evident that the signal state $|G\rangle$ and signal operator $U$ can be prepared efficiently. After that, we can construct the Qubitization intermediate variable $W$ utilizing the signal state $|G\rangle$ and signal oracle $U$ with $\mathcal{O}(1)$ primitive gates. Finally we achieve the Hamiltonian simulation $\langle G|W|G\rangle$, i.e., $\|\langle G|W|G\rangle-e^{-iH^{(l)}t}\|<\varepsilon_{si}$ for time $t$ and error $\varepsilon_{si}$.
Using this technique, the Hamiltonian $H^{(l)}$ can be efficiently simulable in time $\mathcal{O}(\alpha t+\log(1/\varepsilon_{si})/\log\log(1/\varepsilon_{si}))$ \cite{Low2016Hamiltonian}, where $\alpha=\|\sum\limits_{k=1}^{K}w_k\|$ depends on the linear combination of Hamiltonian. For the phase estimation, the propagator $e^{-iH^{(l)}t}$ is enacted with error $\mathcal{O}(1/t)$. Then the Hamiltonian simulation time $t$ determines the error of the phase estimation $\varepsilon_{PE}$. Thus the runtime of the phase estimation step is $\mathcal{O}(1/\varepsilon_{PE})$. Considering the basic computing gates, exp gate and multiplying gate, we can achieve the basic gate in $\mathcal{O}(1)$ time with the help of the Fourier transformation on computational basis technique. Taking into account the measurement of $\Lambda_{X}, \Lambda_{X,Y}$ on each states $|\phi_1\rangle,...,|\phi_m\rangle$, the relative error $\varepsilon_m$ obeys the binomial distribution $m\varepsilon_m=\sqrt{mp_{\lambda}(1-p_{\lambda})}$, where $p_{\lambda}$ is the probability collapsing to the corresponding state. Thus the measurement time $m$ is $\mathcal{O}(1/\varepsilon_m^2)$.

Suppose the training data is denoted as $\{(X^{(1)},Y^{(1)}),...,(X^{(N)},Y^{(N)})\}$, where $(X^{(i)},Y^{(i)})$ is the $i$th observable sequences and corresponding label sequences. The scale of each data block is $|X^{(i)}|=|Y^{(i)}|=n$, and the dimension of the weight parameter $w$ is $K$. The construction of the potential function $E(x,y)$, defined on the probabilistic undirected graph, promises $K\ll n$.  Each component of Boltzmann weight $w$ needs invoking $mN$ times quantum algorithm. Therefore, the overall running time on computing the gradient is
\begin{eqnarray}
\label{Eq:26}
\mathcal{O}(\frac{KN}{\varepsilon_{m}^2}(\frac{\alpha}{\varepsilon_{PE}}+\frac{\log(1/\varepsilon_{si})}{\log\log(1/\varepsilon_{si})})(K+\log\|Q\|)n).
\end{eqnarray}
The parameter $\varepsilon_{PE}$ is the error of phase estimation, $\alpha=\|\sum\limits_{k=1}^{K}w_k\|$ and $\varepsilon_{si}$ is the error of Hamiltonian simulation. Compared to the classical case, which takes $\mathcal{O}(KNn\|Q\|^{n})$ computational overhead, our quantum algorithm achieves exponential acceleration in the training process compared to its classical counterparts.
\subsection{Numerical simulation}
\begin{figure}

  \begin{center}

  \includegraphics[width=0.45\textwidth]{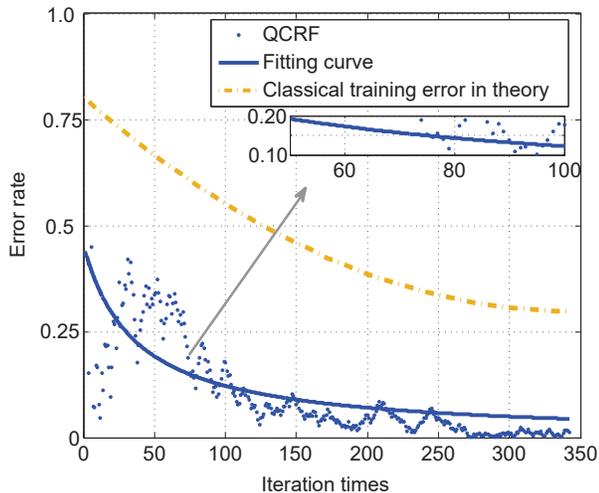}\\
  \caption{The error rate of computing the gradient function}\label{Figure3}
  \end{center}
\end{figure}

Since the quantum model can not be implemented physically by current technology. To evaluate our model and corresponding algorithm, we conduct a representative numerical experiment on a Hamiltonian matrix.\\
\textbf{Datasets Description.} The Hamiltonian matrix $H$ is constructed based on the above section, which encodes $1024$ kinds possible $\sum\limits_i\sum\limits_k w_kf_{ki}$ on its diagonal line. The initial Boltzmann weight is set as $w=(0.17,0.35,0.41,0.52,0.37)$.\\
\textbf{Implementation Details.} We first utilize the fundamental gates to simulate the Hamiltonian $H$ matrix. Then we apply the proposed quantum algorithm onto the Hamiltonian $H$. To be more specific, we freeze the intimal Boltzmann weight $w$ and train our quantum model for $340$ epoches.\\
\textbf{Evaluations on Performance.} We conduct experiments to find the convergence to the error of our quantum algorithm. Experiments results are shown in Fig. \ref{Figure3}. If we recognize each iteration generating one state $|\phi_i\rangle$, the results describe the phenomenon that the downward trend of the estimation error $\varepsilon$ with the iteration times increasing. The blue discrete points are the realistic measurement error rate, and the corresponding fitted error rate curve (i.e., Fitting curve) shows dramatic decline from the iteration times 0 to 340. It is evident that estimation error $\varepsilon$ has tended to stable after 50 iteration, and $\varepsilon$ finally turns to arbitrary small as the increasing of iteration times. The point line indicates the error of classical Gibbs sampling based algorithm. Given an error $\varepsilon$, our quantum algorithm takes fewer steps obtaining $\varepsilon$ compared to the classical method. Thereby, our quantum algorithm shows faster than classical techniques in terms of the rate of convergency.
\section{Discussion: VC dimension}
 According to the classical Bernouli theorem, the relative frequency of an event in a sequence of independent trials converges to the probability of that event. Furthermore, in the learning theory, people prefer to obtain the criteria on the basis of which one could judge whether there is such convergence or not. The VC dimension, proposed by Vapnik et al. \cite{Vapnik1971On}, formulates the conditions for such uniform convergence which do not depend on the distribution properties and furnishes an estimation for the speed of convergence. In briefly, VC dimension evaluates the generalization ability of a learning model. The kernel theory of QCRF lies in $P_Q(y|x)=\frac{Tr(\Lambda_{X,Y}e^{H^{(0)}})}{Tr(\Lambda_{X}e^{H^{(n)}})}$,
which utilizes diagonal Hamiltonian $H^{(l)}$ representing the potential function, nevertheless, off-diagonal Hamiltonian is also permitted. If the instance problem needed, we can add a traverse field onto the system, as a result, the Hamiltonian $H=\sum\limits_{i,k}\vartheta_{k}\sigma_{k,i}^{x}+w_{k}\sigma_{k,i}^z$ turns into an off-diagonal matrix \cite{Wilczek2008Particle}. The construction of extended Pauli X operator $\sigma_{k,i}^{x}$ is similar to $\sigma_{k,i}^z$, which just substitutes $\sigma^x$ to $\sigma^z$ in the $ki$ th position. Compared to the classical CRF model scattering the number of $\Omega(Kn)$ data scales in the whole space, the QCRF model's VC dimension can be calculated as follows. Suppose $w=(w_1,w_2,...,w_K)$ is the Boltzmann weight, and for the input training data $X^{(i)}_t={X^i_1,X^i_2,...,X^i_t}\in R^l$, each $X_j$ is a single word. Now, we analyze the procedure that trains the data from $X_t$ to $X_{t+1}$. Approximate the exponential potential function as a $d$ degree polynomial $e^{x}=\sum\limits_{s=0}^d\frac{x^s}{s!}$, then the data training procedure from $X_t$ to $X_{t+1}$, will increase model's degree. In detail, potential function changes from $\exp(\sum\limits_{i=1}^T\sum\limits_k w_k\sigma_{k,i}^z)$ to $\exp(\sum\limits_{i=1}^{T+1}\sum\limits_k w_k\sigma_{k,i}^z)=\exp(\sum\limits_{i=1}^T\sum\limits_k w_k\sigma_{k,i}^z)\exp(\sum\limits_k w_k\sigma_{k,T+1}^z)$, whose degree is $2d$. After reading the whole input $X(T)={X_1,X_2,...,X_T}$ , the state of any unit in the model can be expressed as a polynomial $P_t,t=1,2,...,T$. The degree of $P_t$ can be estimated as $P_t=2d^t+\sum\limits_{j=1}^{t-1}d^j$. Considering the VC dimension of recurrent structure is bounded by $\Omega(2K\log(8eP_T))$ \cite{Koiran1997Vapnik}, then we obtain the QCRF's VC dimension $\Omega (K2^n)$ when $T=2^n$. It is evident that QCRF model extends this lower bound, which means QCRF can recognize much more samples and features than CRF.

\section{Conclusion}
 In this paper, we have constructed a general QCRF model by introducing four well-defined Hamiltonians and measurements. Meanwhile, in order to train this model, we have also presented an efficient hybrid quantum algorithm to obtain the parameters $w$ in the classical form. Compared to its classical counterpart, the quantum algorithm achieves an exponential speed-up. In addition, numerical simulation results have shown that our QCRF model requires significantly fewer iterations to achieve the same error rate than the classical CRF model. Furthermore, we have also demonstrated that the QCRF model possesses higher VC dimension than the classical CRF model, which means QCRF is equipped with a higher learning ability. We expect our work can inspire more quantum machine learning models and algorithms for handling sequential data.

\end{document}